\documentclass[prx,amsmath,twocolumn,superscriptaddress]{revtex4-1}
\usepackage{graphicx}
\usepackage{amsmath}
\usepackage{amsfonts}
\usepackage{bbm}

\usepackage{color}

\newcommand{\F}{\mathcal{F}}
\newcommand{\be}{\begin{equation}}
\newcommand{\ee}{\end{equation}}

\begin{document}

\title{Scale invariance at the ergodic to many-body localized transition}
\title{Fixed points of Wegner-Wilson flows and many-body localization}
\author{David Pekker}
\affiliation{Department of Physics and Astronomy, University of Pittsburgh, Pittsburgh, PA 15260, USA}
\affiliation{Pittsburgh Quantum Institute, Pittsburgh, PA 15260, USA}
\author{Bryan K. Clark}
\affiliation{Department of Physics, University of Illinois at Urbana Champaign, IL 61801, USA}
\author{Vadim Oganesyan}
\affiliation{Department of Engineering Science and Physics,
College of Staten Island, CUNY, Staten Island, NY 10314, USA}
\affiliation{Physics program, The Graduate Center, CUNY, New York, NY 10016, USA}
\author{Gil Refael}
\affiliation{Department of Physics and the Institute for Quantum Information and Matter,
Caltech, Pasadena, CA 91125, USA}
\begin{abstract}
Many-body localization (MBL) is a 
phase of matter that is characterized by the absence of thermalization. Dynamical generation of a large number of local quantum numbers has been identified as one key characteristic of this phase, quite possibly the microscopic mechanism of breakdown of thermalization and the phase transition itself.
We formulate a robust algorithm, based on Wegner-Wilson flow (WWF) 
renormalization, for computing
these conserved quantities and their interactions.
We present evidence for the existence of distinct fixed point distributions of the latter: a flat ``white noise" distribution in the ergodic phase, a ``$1/f$" law inside the MBL phase, and scale-free distributions in the transition regime.
\end{abstract}

\maketitle

Recent progress on the theory of many-body localization (MBL) demonstrates clearly that the conventional quantum statistical description of 
interacting many-body problems is incomplete.  Concrete analytic~\cite{Basko2006}, numerical~\cite{Oganesyan2007,Znidaric2008,Berkelbach2010,Pal2010} and mathematical~\cite{Imbrie2014,Imbrie2016} results establish the existence and robustness of many-body localized phases in sufficiently strongly disordered and/or low dimensional interacting models 
at finite extensive entropy.  While the understanding of the transition between thermal and MBL phases is only beginning to emerge~\cite{Gopalakrishnan2014,Vosk2015,Potter2015,Yu2015} several distinct new directions of inquiry related to MBL and the fundamental issue of ergodicity in quantum many-body systems have taken shape. 
These include the interplay of MBL with spontaneous symmetry breaking and topological order~\cite{Pekker2014, Huse2013,Chandran2013,Potter2015,Vasseur2015,Vasseur2016}, self-localization (glassiness) in translationally invariant quantum systems~\cite{Schiulaz2014a,Pino2015,Yao2014,VanHorssen2015} and MBL in driven systems~\cite{Lazarides2014,Ponte2014,Khemani2016}.  MBL has also stimulated considerable progress in developing tools for describing excited eigenstates of many-body systems~\cite{Luitz2015, Yu2015, Khemani2016a, Lim2015a, You2016, Monthus2016a}. MBL has been realized in recent experiments~\cite{Schreiber2015,Choi2016} and may also have important implications for ``practical" quantum engineering problems, e.g. quantum computing~\cite{Santos2005,Altshuler2009a,Laumann2015,Yao2015,Khemani2014}.
\begin{figure}[t]
\includegraphics[height=6cm,width=\columnwidth]{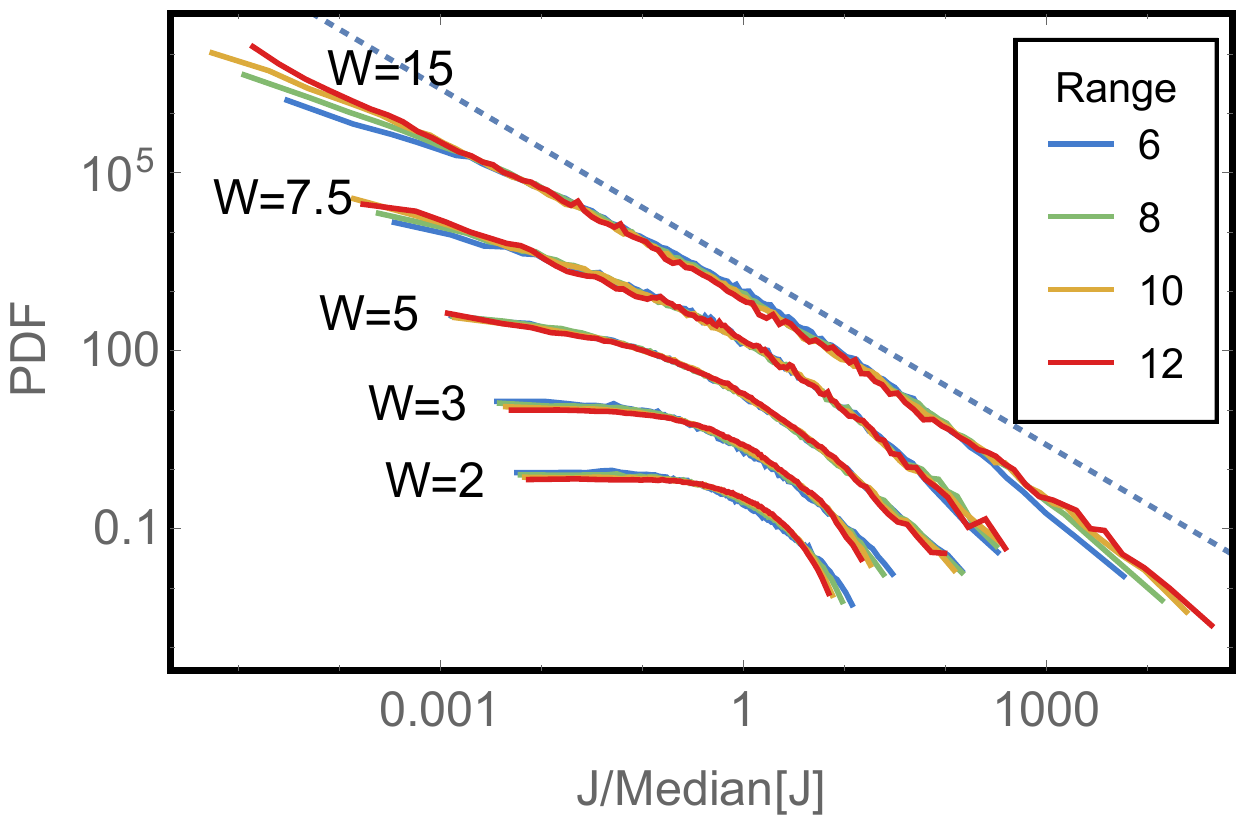}
\vspace{-5mm}
\caption{
Evolution of normalized $\ell$-bit couplings with disorder (vertical offset: $W=2,3,5,7.5,15$) and range (colors, see legend and explanation in text). The straight dotted line corresponds to slope $-1$, i.e. $\sim 1/|J|$ distribution.
\vspace{-5mm}
}
\label{fig:newmoney}
\end{figure}

\begin{figure}[t!]
\includegraphics[width=\columnwidth]{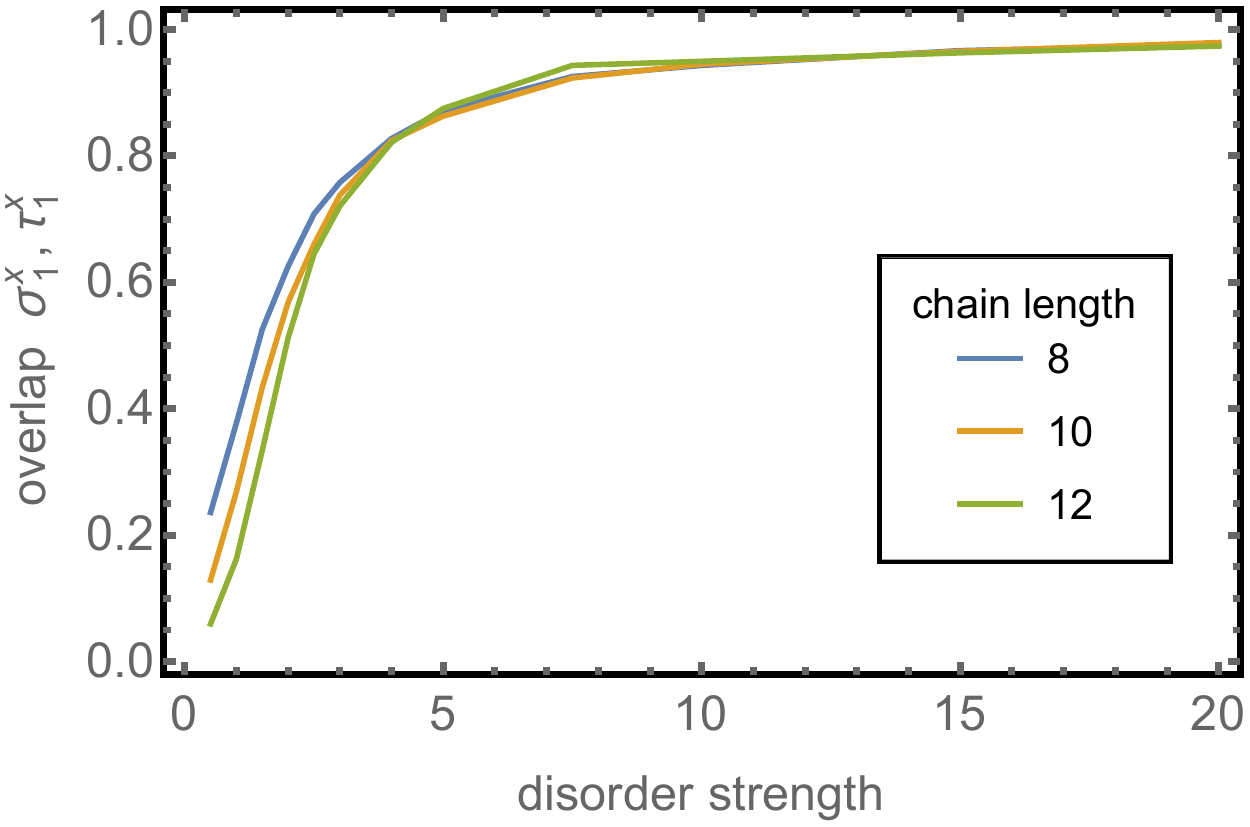}
\caption{
Mean overlap between physical- and $\ell$-bit operators $\sigma^x_1$ and $\tau^x_1$ as a function of disorder strength. The overlap was defined as $\exp \left[ \text{mean}_{\text{disorder}}(\log | \text{Tr} \, \sigma^x_1 \tau^x_1|)\right]$.
}
\label{fig:overlaps}
\end{figure}

One natural route to the breakdown of thermalization is via proliferation of a large number of conserved quasi-local quantities.  The extreme version of such a proposal has gained considerable traction as a model phenomenology~\cite{Huse2014} of the so-called fully-MBL regime, where the entire many-body spectrum is localized. Consider a generic system, e.g. the $n$-site spin $1/2$ random field Heisenberg chain (see Eq. \ref{eq:HH}), which is diagonalized by a unitary matrix $U$.  One can express the diagonal Hamiltonian in terms of $n$ ``elementary" two-level systems ($\ell$-bits) ${\bf \tau}_j=U {\bf \sigma}_j U^+$, such that the entire spectrum is correctly captured by a simple (classical) energy functional on $\tau_j^z$'s only ($\sigma$'s are Pauli matrices representing microscopic spins). Importantly, we expect that for sufficiently strong disorder $\tau$'s are quasi-local~\cite{Ros2015}, i.e. with at most exponential tails, and have finite overlap with the microscopic spin operators $\text{Tr}[{\bf \sigma}_j\cdot{\bf \tau}_j]\neq 0$ in the thermodynamic limit (see Fig. \ref{fig:overlaps}).
This overlap is analogous to the quasiparticle residue in Fermi liquids which allows for direct access to elementary excitations ($\tau$'s in our case) using external probes coupling to microscopic degrees of freedom ($\sigma$'s). Although there is no universally accepted method for constructing $\ell$-bits~\cite{Chandran2014,Pekker2014a,Rademaker2016} as of yet,  one may take finite overlap\cite{Chandran2014} as one ``design criterion".

This work focuses on the nature of interactions among $\ell$-bits. Using a variant of Wegner-Wilson flow transformations, we compute the effective Hamiltonians for short spin chains and detail ``fluctuation"-type aspects, whereby $\ell$-bits $r$ lattice sites apart interact via a \emph{distribution} of effective couplings: we find that the MBL phase tends towards a broad $1/f$ type distribution of couplings, while the ergodic phase follows a self-averaging (``white noise'') law, see Fig. \ref{fig:newmoney}. The transition regime is characterized by scale-invariant ($r$-independent) distributions. Because of the finite overlap with microscopic spin operators these distributions may in principle be inferred experimentally using dynamical protocols. In light of our observations, we can interpret the perturbation theory results of Ref.~\onlinecite{Ros2015} as a possible origin of the broad distributions.

\paragraph*{Methods:} Wegner-Wilson flow is a robust algorithm for constructing (numerical) functional renormalization flow from a given many-body Hamiltonian to its diagonalized form.  In perturbative cases it correctly reproduces results obtained using Feynman diagrams~\cite{Kehrein2006}, however, its true value lies in its non-pertubative nature, rooted in convergence properties for finite systems akin to those of the Jacobi rotation method for exact diagonalization~\cite{Kehrein2006,Quito2016}. Unlike the typical renormalization group schemes, where one ``integrates out" short distance/high energy degrees of freedom to obtain an effective action for the remaining low energy degrees of freedom, WWF works by decoupling degrees of freedom that are separated by large energies without removing any degrees of freedom. The flow generator, $\eta$, is computed\cite{Glazek1993, Glazek1994, Wegner1994, Kehrein2006} by separating the Hamiltonian into diagonal ($H_0$) and off-diagonal ($V$) pieces with respect to a physically motivated (local) basis
\begin{align}
H(\beta)&=H_0(\beta)+V(\beta), \label{eq:wwf1}\\
\eta(\beta)&=[H_0(\beta),V(\beta)],\\
\frac{dU(\beta)}{d\beta}&=\eta(\beta),~\label{eq:wwf6U}\\
\frac{dH(\beta)}{d\beta}&=[H(\beta),\eta(\beta)]~\label{eq:wwf6}.
\end{align}
where $\beta$ is the flow parameter ranging from $0$ to $\infty$. 
Note that we are generally interested not only in $H(\beta)$ but also in $U(\beta)$, the similarity transformation that diagonalizes the problem and from which all other transformed operators may be obtained (see Eq. \ref{eq:wwf6U}).
WWF is a non-linear flow, with the off-diagonal part of $H(\beta)$ flowing to zero continuously and therefore simultaneously reducing the ``size" of $\eta$.  Such flows are robust as they blithely integrate past perturbative resonances and only slow down when the problem is nearly diagonal.  
The initial conditions for the flow are 
\begin{align}
U(\beta=0)&=\mathbbm{1},\\
H(\beta=0)&=H, \label{eq:FlowInitial}
\end{align}
where $H$ is the Hamiltonian we are diagonalizing in the original basis, and $U(\infty)$ and $H(\infty)$ are the quantities of interest.

A few comments are in order before we discuss the results.  First, while the $\ell$-bits are not unique and will depend on the protocol to generate them, our WWF method is entirely deterministic, with an outcome which only depends on the initial basis choice.  The method does bear some resemblance to other iterative diagonalization methods, such as Jacobi rotations or the ``shift method"~\cite{Rademaker2016}, and flow equation method with alternate generators~\cite{Monthus2016}, but it is not equivalent to them. For example, while Jacobi pivots away the largest off-diagonal matrix elements, WWF targets matrix elements connecting the largest energy splittings; alternately, the shift method appears to be organized by the order of number of spin-flips. Also, while other methods are often comprised of discrete steps, WWF is a continuous flow, which may be an important advantage -- in our side-by-side comparison studies (to be published in a separate longer paper) the outcomes of WWF consistently produced ``tighter" unitaries, with less entanglement than those from methods such as bipartite matching~\cite{Pekker2014} and Jacobi iterations.  In fact, we suspect that this may be true generally. 

For the purposes of this Letter, we only compress the structure of $H(\beta)$ and $U(\beta)$ by using sparse representation of these matrices;  we show, however, (see Supplement) that $H(\beta)$ and $U(\beta)$ can be efficiently described by a low bond-dimension matrix-product-operator in the MBL phase and, so, using matrix-product technology could be a fruitful direction to pursue~\cite{Pekker2014b, Chandran2014b, Pollmann2016, Yu2015, Khemani2016mps, Lim2016}. 
In this work, we focus on obtaining and analyzing the ensemble of fixed points $H(\infty)$ and $U(\infty)$ using a straight forward numerical integration of the flow equations~Eqs.\eqref{eq:wwf1}-\eqref{eq:wwf6}. To improve performance, we used several tricks. (1) Numerical integration was performed using Dormand-Prince method [i.e. Runge-Kutta(4,5)]. (2) WWF flow involves a very wide range of RG time scales, spanning from roughly the inverse many-body band-width to the inverse many-body level spacing. To accommodate this wide range of timescales, without resorting to an implicit integration scheme, in the course of integration the very small matrix elements in $H(\beta)$, associated with the short RG timescales, were dropped thus allowing the RG time step to grow as the WWF flow progressed. 
(3) To get the most accurate result, we use WWF to choose the permutation of the columns and rows (and phases) of $U$ to ensure it is local but select the eigenvectors (which in the limit of exact integration must be the same between all approaches) from whichever method gives them to highest precision.

\paragraph*{Model and analysis:}
We consider the random field Heisenberg model
\be
H=\frac{1}{4} \sum_{i} \sigma	_i \cdot \sigma_{i+1} + \frac{1}{2} \sum_i h_{i} \sigma^z_i,
\label{eq:HH}
\ee
on open chains where the $h_{i}$'s are chosen from a uniform distribution $[-w,w]$. In this Letter we focus on the analysis of $H(\infty )$ for (1) a range of chain lengths $L=\{8,10,12\}$, (2) disorder strength spanning the range from $w=0.5$ to $w=20$, and (3) disorder realization (500-1000 disorder realizations were generated for each $L$ and $w$). 

Before we begin the analysis of $H(\infty)$, we examine the possibility of \emph{probing} it using external excitations, e.g. transverse field coupling to $\sigma^x_j$.
To that end we compute and present overlaps between microscopic spin-flip operators $\sigma^x_i$ and $\ell$-bit spin-flip operators $\tau^x_i$ associated with the same site of the chain (see Fig. \ref{fig:overlaps}). In the MBL phase, these overlaps appear to be large and chain length independent. It is likely that these large overlaps persist in the $L\to\infty$ limit. On the other hand, in the ergodic phase, the overlaps are strongly chain length dependent, quickly shrinking as the chain length increases.  The fact that the $\ell$-bit spin-flip operators show a healthy overlap with corresponding microscopic spin-flip operators on the same site implies, among other things, that external time-dependent but local-in-space manipulations can be used to target $\ell$-bit configurations.

\begin{figure}
\includegraphics[width=\columnwidth]{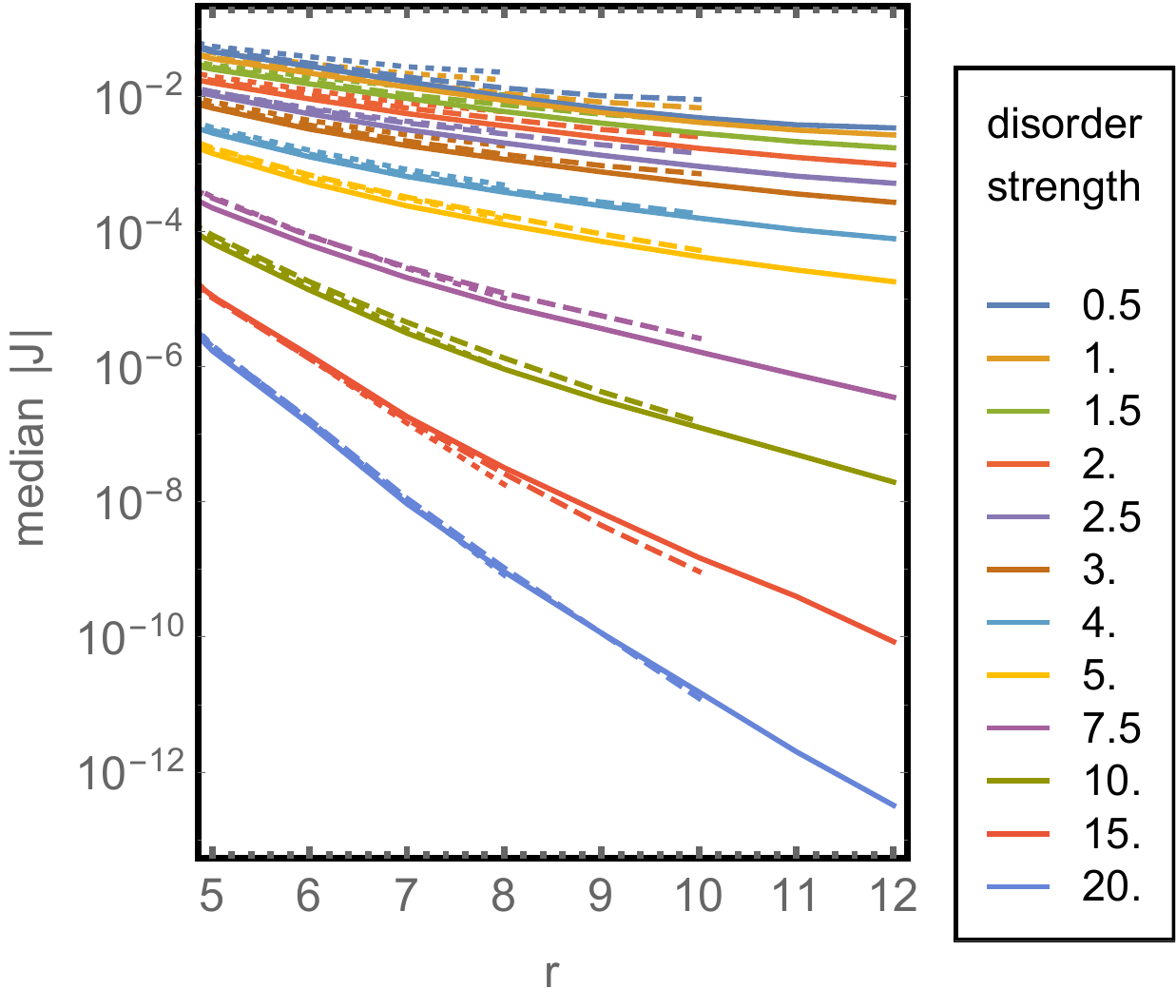}
\caption{
Median $|J|$ as a function of range and disorder strength, for three different chain length ($L=8$ dotted lines; $L=10$ dashed lines, $L=12$ solid lines.
}
\label{fig:medianJ}
\end{figure}

\begin{figure}
\includegraphics[width=\columnwidth]{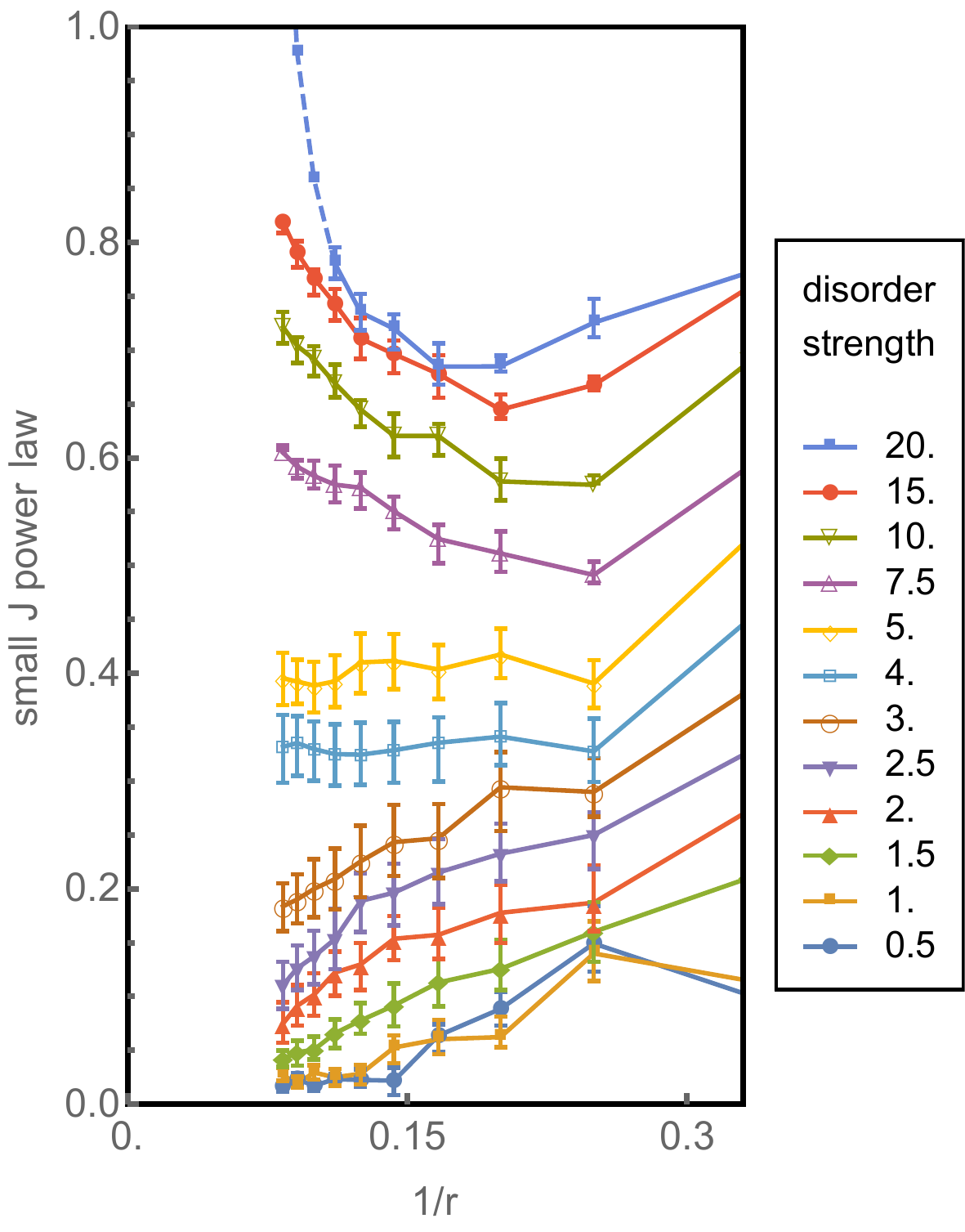}
\caption{
Flows of the end-point Hamiltonian as a function of coarse graining and disorder strength. Specifically, we extract the small $J$ power law from the distributions $\F_{w,r}(J)$ and plot these as a function of $1/r$ for various $w$'s. Observe that flows at weak disorder tend to flat distributions ($J^0$), while flows at strong disorder tend to the $(1/J)$ distribution. In between these two regimes (for $3\lesssim w\lesssim 5$) the power law appears to be independent of the range indicating that the distributions $\F_{w,r}(J)$ are scale-free. The error bars indicate the uncertainty in fitting the small $J$ data to a power law. The dashed line segments for $w=20$ indicate that $\F_{w,r}(J)$ include $J$'s that are below machine precision, and hence an accurate measurement of the power law was not possible.
}
\label{fig:smallJpower}
\end{figure}

We now focus on the analysis of $H(\infty)$
\begin{align}
H(\infty)=E_0 + &\sum_i J_i \tau^z_{i} +  \sum_{i,j} J_{ij} \tau^z_{i}\tau^z_{j} 
\nonumber\\
&\quad\quad+  \sum_{i,j,k} J_{ijk} \tau^z_{i} \tau^z_{j}  \tau^z_{k} + \dots.
\label{eq:Hdiag}
\end{align}
Since WWF preserve all of the information about the many-body problem and because we only have results for few system sizes we need to introduce an additional parameter to elucidate scaling properties of the $\beta\rightarrow\infty$ problem.  As with ordinary criticality, we expect real-space resolution of observables to be a natural direct way to proceed. 
Hence, we introduce the range $r$ which is used to group the coupling constants $J_{i,j,...,k}$ that appear in Eq.~\eqref{eq:Hdiag} by the size of their ``footprint," i.e. the range for the terms $J_{2,5}$, $J_{2,4,5}$, and $J_{4,5,7}$ is $r=4$. For a given $r$ and $w$, we define $\F_{r,w}(J)$ as the distribution function of $|J_{i,j,...,k}|$'s sampled over all disorder realizations. 

We begin by focusing on the gross feature -- the dependence of the typical value of $|J_{i,j,...,k}|$ on the range shown in Fig.~\ref{fig:medianJ}. As expected, there is a strong, approximately exponential decay of median coupling with $r$ in the MBL phase. As the exponential fit is not terribly good, and we do not know an improved functional ansatz (beyond simple exponential) inside the MBL phase, we do not extract an explicit value of the localization length.  Also, perhaps surprisingly, there is an approximately exponential decay of couplings in the ergodic regime. While, at first sight, the behavior in the weak disorder case is surprising, it is indeed consistent with GOE level statistics and hence ergodicity. Specifically, in order to observe GOE statistics for a given range $r$ the typical value of $|J_{i,j,...,k}|$ must exceed the level spacing $r 2^{-r}$. This condition is indeed satisfied for our data in the weak disorder regime $w \lesssim 4$.

We now turn to the full counting statistics of $J$'s which appears to show a much clearer ``flow" with $r$ than the median $J$ itself, see Fig. \ref{fig:newmoney}. There are three clearly distinguishable regimes: (i) the couplings ``flow" to the $1/|J|$ law everywhere in the MBL phase; (ii) the couplings ``flow" to the approximately constant distributions (possibly gaussian?) in the ergodic phase; (iii) the couplings do not flow in the intermediate, ``critical" regime. The full distribution functions $\F_{r,w}({\cal J})$ appear to form a one-parameter family $\F_{f(r,w)}({\cal J})$ (see Supplement). At present we do not have sufficient resolution to opine on whether the critical point is unique or there is a critical ``phase" separating the ergodic and MBL phases.  

Focusing on the small $|J|$'s we can recast these qualitative observations into a quantitative fit to power-law behavior $\F_{r,w}({\cal J})\propto {\cal J}^{-\alpha_{r,w}}$ for the small ${\cal J}$ part of the curve. We plot $\alpha_{r,w}$ as a function of $1/r$ in Fig.~\ref{fig:smallJpower}. 
As already foretold visually in Fig. \ref{fig:newmoney} there is a flow (as $r\to \infty$) in $\alpha$ towards respectively white noise and $1/f$ laws below and above the critical regime residing near $3\lesssim w \lesssim 5$.

\paragraph*{Summary and outlook:}
We have applied a numerical implementation of the Wegner-Wilson flow renormalization group to random field Heisenberg chains. 
The properties of the fixed point (diagonal) Hamiltonians and corresponding unitaries are consistent with the phenomenology of fully MBL matter~\cite{Huse2014} when disorder is sufficiently strong. We have investigated the range-dependence of the end-point diagonal Hamiltonians produced by Wegner-Wilson flow. We found robust flow towards broad $1/f$-type distributions in the MBL phase and narrow white-noise-like distributions in the ergodic phase. At intermediate disorder, we found what appears to be a scale-free critical point or critical phase that demarcates the boundary between the ergodic and the MBL phases. To quantify these trends, we analyzed power laws in the small-$J$ tails of the distribution. The dependence of the extracted power laws on range revealed bi-furcating ``flows" that seem to be an essentially universal feature of the MBL transition.

The successful numerical application of the Wegner-Wilson flow approach to this important problem points to the possibility of an analytical treatment of the MBL transition using the same methodology.

\paragraph*{Acknowledgements:}
We are grateful to A. Scardicchio, S. Kehrein, E. Kapit, A. Chandran, D. Huse, V. Khemani, B. Altshuler, L. Rademaker, M. Ortu\~{n}o, X. Yu for stimulating discussions. 

D.P. acknowledges support from the Charles E. Kaufman foundation. BKC was supported by SciDAC-DOE grant DE-FG02-12ER46875. GR is grateful for support from the Institute of Quantum Information and Matter, an NSF Frontier center funded by the Gordon and Betty Moore Foundation, and the Packard foundation. V. O. acknowledges support from the NSF DMR Grants Nos. 0955714 and 1508538. Parts of this work were performed at the Aspen Center for Physics, which is supported by National Science Foundation grant PHY-1066293 (D.P., B.K.C., G.R., and V.O.), ICTP Trieste (V.O.), and KITP Santa Barbara, which is supported by National Science Foundation under Grant No. NSF PHY11-25915 (D.P, G.R., and V.O.). The authors thank all three centers for the hospitality offered to us. This research is part of the Blue Waters sustained petascale computing project, which is supported by the National Science Foundation (awards OCI-0725070 and ACI-1238993) and the State of Illinois. Blue Waters is a joint effort of the University of Illinois at Urbana Champaign and its National Center for Supercomputing Applications.

\bibliography{library,notes}
\end{document}